# 1-D Convolutional Graph Convolutional Networks for Fault Detection in Distributed Energy Systems

Bang L.H. Nguyen[†§], Tuyen Vu[†], Thai-Thanh Nguyen[‡], Mayank Panwar[§], Rob Hovsapian[§],
[†]*ECE Department, Clarkson University*, Potsdam, NY, USA
[‡]*New York Power Authority*, White Plains, NY, USA
[§]*National Renewable Energy Research Laboratory*, Golden, CO, USA
bang.nguyen@nrel.gov, tvu@clarkson.edu, thaithanh.nguyen@nypa.gov,
mayank.panwar@nrel.gov, rob.hovsapian@nrel.gov

*Abstract*— **This paper presents a 1-D convolutional graph neural network for fault detection in microgrids. The combination of 1-D convolutional neural networks (1D-CNN) and graph convolutional networks (GCN) helps extract both spatial-temporal correlations from the voltage measurements in microgrids. The fault detection scheme includes fault event detection, fault type and phase classification, and fault location. There are five neural network model training to handle these tasks. Transfer learning and fine-tuning are applied to reduce training efforts. The combined recurrent graph convolutional neural networks (1D-CGCN) is compared with the traditional ANN structure on the Potsdam 13-bus microgrid dataset. The achievable accuracy of 99.27%, 98.1%, 98.75%, and 95.6% for fault detection, fault type classification, fault phase identification, and fault location respectively.**

## I. Introduction

Fault diagnostic plays a key role to determine the strategy of how to isolate and restore power systems, especially under the growing integration of distributed energy resources. This protection and restoration strategy ensures the system's resiliency and reliability [1], [2]. In inverter-based distributed energy resources, the traditional relay protection may become ineffective due to the small fault current [3], [4], [5]. Moreover, to effectively and accurately isolate faults and restore normal operation, one requires the information of fault event, fault type, fault phase, and fault location [6], [7]. The correct information about faults significantly enhances the protection and restoration and also saves time and cost of utilities [8], [9], [10].

The fault diagnostic schemes existing in literature [11]–[24] can be loosely divided into model-based and data-driven methods. The measurements are voltage and current with different sampling rates from digital relays, phasor measurement units (PMU), or advance metering infrastructure (AMI) [25]. Model-based techniques try to compute quantitative metrics that distinguish fault data from normal measurements. A comparison between pre-fault data and fault data is usually evaluated for fault detection [26], [27]. There are many analytical approaches are applied such as evaluating the negative and positive sequences of current [26], assessing the sequential voltage and current components [28], monitoring the transient of current [27], computing the Teager-Kaiser energy [29], analyzing the principal components and fault signatures [30], and state estimation using mathematical morphology and recursive least square [31].

Data-driven and machine learning-based approaches try to derive a fault detection model using statistical information from the measurement data. There are many popular machine learning classifiers have been applied to detect faults such as decision tree (DT) [4], random forest (RF) [32], k-nearest neighbor (k-NN), support vector machine (SVM), and Naïve Bayes [33]. Model-based and machine learning can be combined in the way that model-based techniques do the feature extraction and machine learning do the classification. In [10], discrete wavelet transform is applied before the classification process. The maximal overlap discrete wavelet transform and extreme gradient boost algorithm are employed in [34]. Pure neural network structures are employed frequently such as Taguchi-based artificial neural networks [35], and gated-recurrent-unit deep neural networks [36].

Most existing works analyze the current measurements on the line the fault occurs. There is some fault detection scheme using PMU and pseudo-measurements [37], [38], [39]. Similarly, the machine learning techniques of SVM, k-NN, DT algorithms [39], convolutional neural networks (CNN) [40], [41], semi-supervised [42], and GCN [43] are implemented to detect faults. However, in these works, there is a research gap in fault type, fault phase classification, and fault location on mesh-topology power systems.

This paper presents a combination of 1-D convolutional NN and graph learning on voltage measurement data to detect fault events, classify fault type and phase, and locate the nearest bus where the fault occurs. The paper provides a unique contribution owing to the following bullet points.
- The data input includes voltage measurements in time series from PMU, AMI, or smart meters. The time synchronization for phasors is not necessary.
- The combination of 1-D CNN and GCN can extract

- both spatial and temporal correlation in the measurement data.
- The fault event detection, fault type, phase classification, and fault location are all resolved.

The remaining parts are organized as follows. Section II presents the Potsdam microgrids and the graph data collection procedure. In Section III, the combination of 1-D CNN and GCN is described. The training, transfer learning, and fine-tuning processes are also expressed. The results are discussed in Section V. Section VI concludes the paper.

## II. GRAPH DATASET OF POTSDAM MICROGRID

The power distribution networks can be defined as an undirected graph $\mathcal{G} = (\mathcal{V}, \mathcal{E}, \mathcal{A})$, where $\mathcal{V}$ denotes the set of vertices, $|\mathcal{V}| = N$, each vertex in the graph represents a node (bus) in the distribution network, $X = \{X_1, X_2, \ldots X_N\}$ is the tuple of node features, $\mathcal{E}$ denotes the set of edges, $|\mathcal{E}| = M$, each edge represents a branch connecting two buses, $E = \{E_1, E_2, \ldots E_M\}$ is the tuple of edge feature, and $\mathcal{A} \in \mathbb{R}^{N \times N}$ denotes the adjacency matrix of the distribution network. The input data for graph learning are the node features $X_{i=1\ldots N}$, and the edge features $E_{i=1\ldots M}$. Some papers also consider the edge features and the attributes for each graph data ($u$) [44]; however, in this paper, we only consider the node features on a graph. The temporal graph dataset is constructed by the ordered set of graph, node feature matrix, and label vector tuples [45] $\mathcal{D} = \{(\mathcal{G}^1, X^1, y^1), (\mathcal{G}^2, X^2, y^2), \ldots (\mathcal{G}^I, X^I, y^I)\}$, where the vertex sets is unchanged $\mathcal{V}^i = \mathcal{V}, \forall i \in \{1, \ldots, I\}$, $i$ is the graph data index. The node feature matrices $X^i \in \mathbb{R}^{N \times d \times K}$ have 3 dimensions as follows: the number of nodes $|\mathcal{V}| = N$, the number of features in each node $d$, and the time interval $K$. The label vector includes 3 labels of the distribution network graph over the time interval $K$, $y^i = \{y_{type}, y_{phase}, y_{loc}\}$, where $y_{loc}$ is the node index where the fault occurs. The node feature matrix $X^i = \{X_1, X_2, \ldots X_N\}$ contains the bus voltages of all measured buses. In the bus without voltage measured, the node features are filled with zeros. The node feature in node $i$ is shown in the form of

$$X_i = \begin{bmatrix} V_{a,1} & V_{a,2} & \cdots & V_{a,K} \\ V_{b,1} & V_{b,2} & \cdots & V_{b,K} \\ V_{c,1} & V_{c,2} & \cdots & V_{c,K} \end{bmatrix}^T, \quad (1)$$

where $K$ is the length of the evaluation period.

Specifically, considering the Potsdam microgrid shown in Fig. 1, we have a graph of 13 nodes and 13 edges. There are 5 inverter-based generators (IBG) with a primary droop control strategy [46] and a secondary PI controller for frequency and average voltage regulation [47] in the islanded mode. The voltage level is 13.2 kV line-line at 60 Hz. The loads and IBGS have parameters following those of [48]. The voltage measurements are placed in the buses marked with a blue square; the data sampling frequency is 1 kHz. The data is collected via real-time simulation using Opal-RT. Load changes are set randomly between 30-130% of the nominal load profile. Faults are set at each bus in turn with the fault type of AG, BG, CG, AB, BC, CA, ABG, BCG, CAG, ABC, and ABCG and fault resistance of 0.1, 1, and 10 Ω. The raw data are collected as one second windows and then are trimmed into 20 ms of 20 samples which cover about 1.2 cycles of 60 Hz as shown in Fig. 2. Thereafter, 55,770 graph data of 20-ms windows for the fault cases and 8,580 graph data of non-fault cases with random load changes are gathered as the train set. We also select 8,580 fault and 1,420 non-fault cases for the test set. Table I summarizes these configurations for fault cases and load changes data generation.

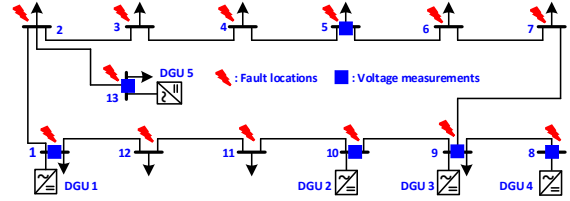

Fig. 1. 13-bus Potsdam microgrid system diagram with fault locations and voltage measurements on buses 1, 5, 8, 9, 10, and 13.

TABLE I. POTSDAM MICROGRID DATASET

| Parameters | Configuration | Count |
|---|---|---|
| Fault type | AG, BG, CG, AB, BC, CA, ABG, BCG, CAG, ABC, ABCG | 11 |
| Fault resistance | 0.1, 1, 10 (Ω) | 3 |
| Fault location | **Buses**: 1, 2, 3, 4, 5, 6, 7, 8, 9, 10, 11, 12, 13. | 13 |
| Load scenario | randomly | 150 |
| **Total fault cases: 64,350** | **Train: 55,770** | **Test: 8,580** |
| **Total load change cases: 10,000** | **Train: 8,580** | **Test: 1,420** |
| **Train-set: 64,350 samples** | **Test-set: 10,000 samples** | |

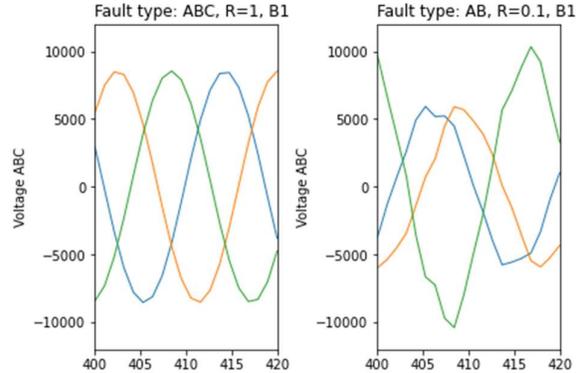

Fig. 2. Voltage waveform in phases A, B, C at bus 1 with ABC and AB faults and fault resistance 1 and 0.1 Ω occurs at bus 1, respectively, in the Potsdam microgrid.

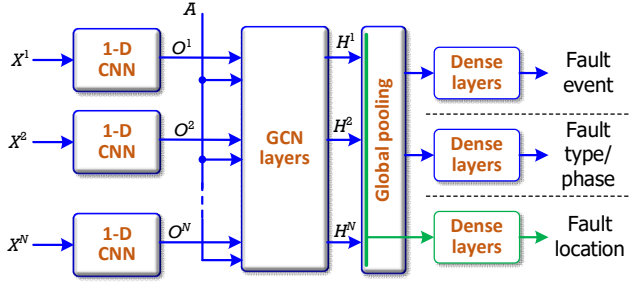

Fig. 3. Proposed temporal 1D-CGCN structure for fault detection.

## III. 1-D CONVOLUTIONAL GRAPH NEURAL NETWORKS MODELS FOR FAULT DETECTION.

In this paper, we utilize the 1-D CNN to extract the temporal correlation in time series data of voltage measurement in each bus. Thereafter, the GCN layers are used to generalize the spatial correlation on graph of the Potsdam microgrid. The proposed temporal 1D-CGCN structure for fault detection is depicted in Fig. 3. This structure includes the 1D-CNN layer and the GCN layer for spatial-temporal feature extraction. Firstly, the 1D-CNN layer is utilized to extract the time-series feature from the voltage measurement of each node. Thereafter, the GCN layer is used to derive the spatial correlation between the bus voltages over all buses on the distribution system. The global pooling operation concentrates all hidden features from nodes and finally, the dense layers are trained to classify the fault type and fault phase. The fault location is performed based on all hidden features from all the nodes. The formulation of 1-D CNN and GCN layers is presented as follows.

*1-D Convolutional Neural Network*:

The 1D CNN layer [49] is expressed as

$$o_k^l = \sigma\left[\sum_{i=1}^{N_{l-1}} Conv1D(\omega_{ik}^{l-1}, o_i^{l-1}) + b_k^l\right], \quad (2)$$

where $o_i^0 = X_i$ is the input feature, $o_i^{l-1}$ is the output of 1-D CNN layer $l$, $\omega_{ik}^{l-1}$ are the trainable weights at layer $l-1$, $o_i^l$ is the output of 1-D CNN layer $l$, $Conv1D$ is the valid cross-correlation operator.

*Graph Convolutional Network*:

The node feature at each time index is processed by the GCN layers [50], which can be expressed as

$$H_{(l+1)}^i = \sigma\left(\widetilde{D}^{-\frac{1}{2}}\widetilde{A}\widetilde{D}^{-\frac{1}{2}}H_{(l)}^i W_{(l)}\right), \quad (3)$$

where $\widetilde{A} = A + I_N$ is the adjacent matrix with self-connection, $I_N$ is the identity matrix, $\widetilde{D}$ is the agree matrix from $\widetilde{A}$ with $\widetilde{D}_{ii} = \sum_i \widetilde{A}_{ij}$ and $\widetilde{D}_{ij} = 0$, $U_{(l)}^i$ is the output of layer $l$, $H_{(0)}^i = O^i$, $W_{(l)}$ is the weight matrix of layer $l$, $\sigma(\cdot)$ is a nonlinear activation function. This graph propagation formula can be derived as a first-order approximation of localized spectral filers [44].

TABLE II.  COMPARISONS OF NEURAL NETWORK STRUCTURES

| ANN | | 1D-CGCN | |
|---|---|---|---|
| **Shared feature extraction layers** | | | |
| Input | [780] | Input | [13×3×20] |
| Dense | [512] | 1D-CNN | [13×3×5] |
| Dense | [128] | GCN | [13×8] |
| **Fault event binary classification – Dense layers** | | | |
| Dense | [32] | Dense | [16] |
| Dense | [1] | Dense | [1] |
| **Fault location – Dense layers** | | | |
| Dense | [64] | Dense | [13×8] |
| Dense | [13] | Dense | [13] |
| **Fault type classification– Dense layers** | | | |
| Dense | [64] | Dense | [32] |
| Dense | [6] | Dense | [6] |
| **Fault phase classification– Dense layers** | | | |
| Dense | [64] | Dense | [32] |
| Dense | [3] | Dense | [3] |

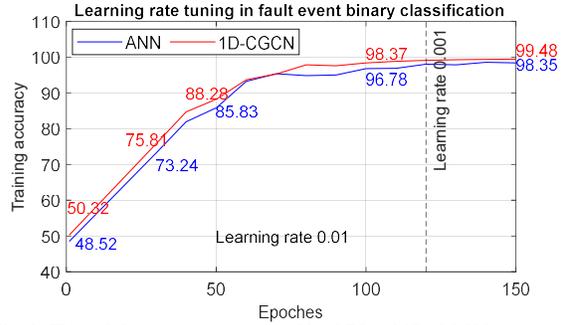

Fig. 4. The training accuracy curves with ANN and 1D-CGCN structures under the change of learning rate from 0.01 to 0.001 at epoch 120.

The detail structures of ANN and 1D-CGCN are compared in Table II, where we have shared layer for feature extraction and dense layers for classification models or classifiers. Reshaping and flattening operations are applied appropriately to condition the dimension compatibility between layers. There are 4 classifiers for fault event detection, fault location, fault type classification, and fault phase identification. The outputs of fault event detection are fault and no-fault. The fault types are classified into six types included **1)** no-fault (**NF**, **2)** single-phase-to-ground (**LG**), **3)** two-phase (**LL**), **4)** two-phase-to-ground (**LLG**), **5)** three-phase (**3L**), and **6)** three-phase-to-ground (**3LG**). Therefore, $y_{type} \in \mathbb{B}^{1\times 6}$ with the $i$-th element of $y_{type}$: $y_{type}[i] = 1$ indicates the $i$-th fault category occurred while all other $y_{type}[i] = 0$. The fault phases are determined by $y_{phase} \in \mathbb{B}^{1\times 3}$, where $y_{phase}[i] = 1$ indicating the fault occurs in phase $A, B, C$, or $AB, BC, CA$ when the fault types are asymmetrical i.e. LG, LL, and LLG, respectively. The fault location is indicated by $y_i = 1$, where $i = 1, 2, 3, ... N$ if the fault occurs in the $i$-th bus, otherwise $y_i = 0$. The fault location detection is performed at node-level classification, where the faulty bus is labeled as 1 and the non-fault bus is labeled as 0.

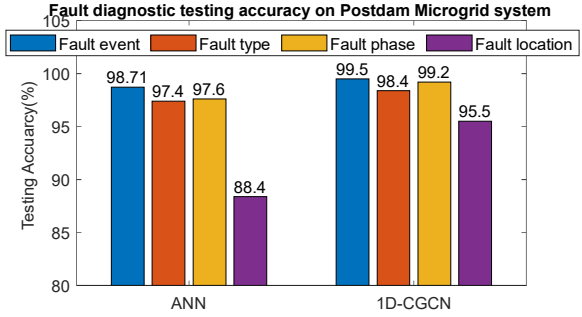

Fig. 5. Fault detection accuracy of Potsdam Microgrid system using proposed 1D-CGCN in comparison with ANN structure.

Fig. 6. Confusion matrix for fault type classification using 1D-CGCN of Potsdam microgrid test set.

Fig. 7. Confusion matrix for fault phase A, B, and C classification using 1D-CGCN of Potsdam microgrid test set.

Fig. 8. Confusion matrix for fault phase AB, BC, and CA classification using 1D-CGCN of Potsdam microgrid test set.

The graph dataset is trained with Adam optimizer and cross-entropy losses. The random dropout of 10% is added in dense layers to reduce overfitting. The learning rate is started at 0.01 and then is reduced to 0.001 at epoch 120 as shown in Fig. 4. As can be seen, the training accuracies of ANN and 1D-CGCN achieve 99.48% and 98.35%, respectively.

After training the fault event classification 120 epochs, the shared feature extraction layers are transferred into 3 other models. Specific dense layers are added to train again for fault location, fault type, and phase classification. Firstly, transfer learning is performed since we freeze the transferred layers and only do training for the additional dense layers. After 120 epochs, we unfroze those transferred layers and train again the entire models with 0.001 learning rate for the fine-tuning process.

## IV. RESULTS AND DISCUSSION

The training and test results are collected on a personal computer with Intel Core i7-8700, 32 GHz, 32 GB RAM, and NVIDIA GTX 1080 GPU. The machine learning framework is Pytorch with Pytorch-geometric library for graph learning [51].

The fault detection accuracies of ANN and the proposed 1D-CGCN are compared in Fig. 5. As can be seen, for the fault event detection, ANN achieves 98.71% while 1D-CGCN can achieve 99.5%. For the fault type classification, 1D-CGCN have 1% higher than ANN since the two structures achieve 97.4% and 98.4% respectively. The 1D-CGCN is outperformed in fault phase identification with 99.2% compared to 97.6% of the ANN. Similarly, the 95.5% accuracy with 1D-CGCN in fault location compared to only 88.4% of ANN.

The detailed confusion matrix of fault type classification is shown in Fig. 6. The detailed confusion matrices of fault phase identification are shown in Figs. 7 and 8. There are 780 graph data for each line-ground (LG) fault and 1560 graph data for each line-line and double line-ground (LL and LLG) fault. The values in those confusion matrices are consistent with the testing accuracy in Fig. 5. Those results prove the high performance of the proposed fault detection models using 1D-CGCN.

## V. CONCLUSION

In this paper, we propose a combination of 1D-CNN and GCN named 1D-CGCN for fault detection in distributed energy systems. The voltage measurements are inputs of the fault detection models. The detection models handle fault event detection, fault type and phase classification, and fault location. The real-time simulation graph data from the Potsdam microgrid using Opal-RT are collected and trained for the models. Transfer learning and fine-tuning techniques are applied to reduce training efforts. The performance of 1D-CGCN is compared with the traditional ANN to prove its superiority. The detailed confusion matrices of the classification tasks are shown for validation.

Although the proposed 1D-CGCN can achieve high accuracies, however, the effects of measurement noises and the lack of measurement data are not considered. Those issues would be tackled in future work.

ACKNOWLEDGMENT

The information, data, or work presented herein was funded in part by the 1) U.S. Office of Naval Research under award number N000142212239, and 2) National Science Foundation (NSF-AMPS) under award number CON0002619.